\documentclass[twocolumn]{jpsj3}
\usepackage{txfonts}
\usepackage{bm}

\title{Spin Susceptibility in Non-Centrosymmetric Superconductors with Topological Transition of Fermi Surfaces}

\author{Daisuke MARUYAMA$^{1}$\thanks{E-mail address: marudai@phys.sc.niigata-u.ac.jp} and Youichi YANASE$^{1,2}$ 
}
\inst{
$^{1}$Graduate School of Science and Technology, Niigata University, Niigata 950-2181, Japan \\ $^{2}$Department of Physics, Niigata University, Niigata
950-2181, Japan } 

\abst{The non-centrosymmetric superconductors Li$_2$Pd$_3$B and Li$_2$Pt$_3$B show 
different superconducting properties despite having the same crystal symmetry. 
Motivated by experimental results, we investigate the spin susceptibility of 
non-centrosymmetric superconductors accompanied by the topological transition 
of Fermi surfaces due to antisymmetric spin-orbit coupling, which is 
indicated by the first-principles band structure calculation for Li$_2$Pt$_3$B. 
We study three types of topological transition, namely, (A) the disappearance of the Fermi surface, 
(B) crossing the Dirac point, and (C) crossing the saddle point van-Hove singularity. 
The spin susceptibility in the superconducting state is increased by 
the topological transitions (A) and (C), while it is decreased by (B). 
We discuss the unusual magnetic properties observed in Li$_2$Pt$_3$B on the basis of these results. 
}

\kword{superconductivity without inversion symmetry, topological transition of Fermi surfaces, spin susceptibility}

\begin{document}

\maketitle

\section{Introduction}

Recently, superconductors lacking inversion symmetry in the crystal structure 
have been attracting much attention. 
The antisymmetric spin-orbit coupling induced by the broken inversion symmetry
leads to the spin-splitting of the Fermi surface and 
gives rise to unique superconducting properties, such as the parity mixing of 
Cooper pairs~\cite{NCSC}.

Among many non-centrosymmetric superconductors, 
the perovskite-like cubic compounds Li$_2$Pd$_3$B~\cite{PhysRevLett.93.247004}
and Li$_2$Pt$_3$B~\cite{JPSJ.74.1014} show particularly intriguing properties. 
The superconducting properties are different between these two compounds 
in spite of having the same crystal symmetry.~\cite{PhysRevLett.93.247004,JPSJ.74.1014,PhysRevB.73.214528,PhysRevLett.97.017006,PhysRevLett.98.047002,PhysRevB.76.104506,
PhysRevB.71.092507,PhysRevB.71.220505,PhysRevB.72.104506,PhysRevB.84.054521,PhysRevB.86.220502}
The order parameter is fully gapped in Li$_2$Pd$_3$B, while it has line nodes in Li$_2$Pt$_3$B.~\cite{PhysRevB.73.214528,PhysRevLett.97.017006,PhysRevLett.98.047002,PhysRevB.76.104506,PhysRevB.71.220505,PhysRevB.86.220502}
The NMR Knight shift of Li$_2$Pd$_3$B is decreased across the superconducting transition temperature $T_{\rm c}$, while that of Li$_2$Pt$_3$B is mostly unaffected in the superconducting state.~\cite{PhysRevLett.98.047002,PhysRevB.86.220502} 
Although these behaviors of Li$_2$Pd$_3$B indicate the conventional s-wave superconductivity 
admixed with the spin triplet p-wave one owing to the antisymmetric spin-orbit coupling, 
experimental results of Li$_2$Pt$_3$B are incompatible with the canonical theory of 
non-centrosymmetric superconductivity.~\cite{NCSC} 

According to the weak coupling theory neglecting the correlation effects, 
the spin susceptibility in the cubic non-centrosymmetric superconductor should 
be reduced to $2/3$ of the normal state value at $T=0$.~\cite{JPSJ.76.034712,PhysRevB.76.094516} 
On the other hand, the Knight shift measurement of Li$_2$Pt$_3$B did not show such a decrease in 
spin susceptibility.~\cite{PhysRevLett.98.047002,PhysRevB.86.220502} 
The spin triplet superconducting state has been proposed for Li$_2$Pt$_3$B 
on the basis of this experimental result;~\cite{PhysRevLett.98.047002,PhysRevB.86.220502} 
however, the spin susceptibility at low magnetic 
fields $\mu_{\rm B}H \ll k_{\rm B} T_{\rm c}$ is independent of the symmetry of the order 
parameter.~\cite{JPSJ.76.124709}
Indeed, the pairing states indicated by theoretical 
studies~\cite{PhysRevB.86.134526,JPSJ.82.024703} are incompatible with the Knight shift measurement 
of Li$_2$Pt$_3$B. 
Although the electron correlation effect may enhance the spin susceptibility 
in the superconducting state,~\cite{JPSJ.76.034712} such enhancement is unlikely to occur
in Li$_2$Pd$_3$B and Li$_2$Pt$_3$B in which the correlation effect is negligible.~\cite{PhysRevB.71.092507,PhysRevB.72.104506} 
The influence of magnetic order has been pointed out for the heavy fermion superconductor 
CePt$_3$Si;~\cite{JPSJ.76.043712,JPSJ.77.124711}
however, the magnetic order does not occur in Li$_2$Pt$_3$B. 
Thus, the superconducting state of Li$_2$Pt$_3$B remains controversial, while 
Li$_2$Pd$_3$B is a ``conventional'' non-centrosymmetric superconductor.

For the difference between Li$_2$Pt$_3$B and Li$_2$Pd$_3$B, a substantial enhancement of 
antisymmetric spin-orbit coupling has been pointed out for Li$_2$Pt$_3$B.~\cite{PhysRevB.72.174505} 
The increase in atomic {\it LS}-coupling on Pt ions as well as the deformation of 
crystal structure~\cite{PhysRevB.86.220502} significantly increases the antisymmetric spin-orbit 
coupling of Li$_2$Pt$_3$B. According to the first-principles band structure calculation, 
this enhancement of antisymmetric spin-orbit coupling is accompanied by 
the topological transition of the Fermi surfaces (FS topological transition).~\cite{Shishidou} 
The Fermi surfaces of Li$_2$Pd$_3$B consist of several pairs of spin-split Fermi surfaces. 
On the other hand, the counterpart of some pairs vanishes in Li$_2$Pt$_3$B. 
According to recent studies of the crystal structure of solid solution Li$_2$(Pd$_{1-x}$Pt$_x$)$_3$B, 
the structural deformation occurs at approximately $x \sim 0.8$,~\cite{PhysRevB.86.220502} which is probably 
accompanied by the FS topological transition.

\begin{figure}[tbp]
  \begin{center}
    \includegraphics[scale=0.35]{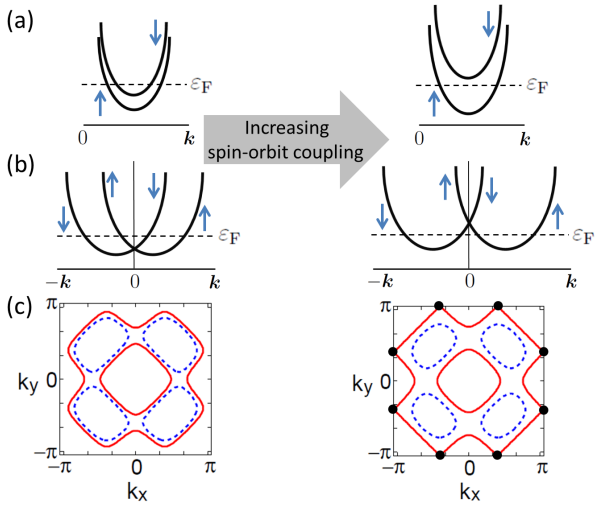}
  \end{center}
  \caption{(Color online)
    Schematic figure of FS topological transitions. 
    (a) Type (A): One of the split Fermi surfaces vanishes around the top or bottom of the band. 
    (b) Type (B): The Fermi level crosses the Dirac point. 
    In figures (a) and (b), the thick solid lines show the two spin-split bands and 
    the arrows show the spin of each band. 
    (c) Type (C): A Fermi surface crosses the van-Hove singularities (closed circles). 
    The dotted and solid lines show the Fermi surfaces of the $\varepsilon_+$ band and the $\varepsilon_-$ band, respectively.
  }
  \label{exfs}
\end{figure}

The purpose of this study is to clarify the effect of FS topological transition 
due to the antisymmetric spin-orbit coupling on the superconducting state.
We here study the roles of three types of FS topological transition 
on the spin susceptibility in the superconducting state. 
In type (A), one of the spin-split Fermi surfaces vanishes owing to the 
substantial increase in the spin-orbit coupling [see Fig.~\ref{exfs}(a)]. 
In type (B), the Fermi surface crosses the Dirac point with increasing spin-orbit coupling, 
as shown in Fig.~\ref{exfs}(b). 
Finally, in type (C), one of the spin-split Fermi surfaces crosses a van-Hove singularity, 
as shown in Fig.~\ref{exfs}(c). 
Our analysis is based on a single-band model, which cannot reproduce the electronic structure of 
Li$_2$(Pd$_{1-x}$Pt$_x$)$_3$B. However, some effects of FS topological transition are  
independent of the band structure, as shown below. We expect that the following results would be a 
key to resolve unsettled issues of Li$_2$(Pd$_{1-x}$Pt$_x$)$_3$B. Our results are also 
applicable to other non-centrosymmetric superconductors with a large spin-orbit coupling. 
We introduce the model Hamiltonian in Sect.~2 and show the numerical results of 
spin susceptibility in Sect.~3. Some remarks are given in Sect.~4.

\section{Model}

We adopt the following single-band Hamiltonian; 
\begin{eqnarray}
\label{model}
H&=&\sum_{\mib{k},s}\varepsilon(\mib{k})c^{\dag}_{\mib{k}s}c_{\mib{k}s}
+\alpha\sum_{\mib{k},s,s'}\mib{g}(\mib{k})\cdot\mib{\sigma}_{ss'}c^{\dag}_{\mib{k}s}c_{\mib{k}s'} \nonumber \\
&+&\frac{1}{2}\sum_{\mib{k},s,s'}\Bigl[\Delta_{ss'}(\mib{k})c^{\dag}_{\mib{k}s}c^{\dag}_{-\mib{k}s'}+\rm{h.c.}\Bigr],
\end{eqnarray}
where $c_{\mib{k}s}$ ($c^{\dag}_{\mib{k}s}$) is the annihilation (creation) operator for an electron 
with a momentum $\mib{k}$ and a spin $s$.  
The dispersion relation $\varepsilon(\mib{k})$ is assumed so that the FS topological transition 
occurs with increasing antisymmetric spin-orbit coupling.
The chemical potential $\mu$ is involved in the dispersion relation and determined so that 
the electron density per site is $n$.
The second term describes the antisymmetric spin-orbit coupling, which preserves the 
time reversal symmetry for the antisymmetric g-vector $\mib{g}(-\mib{k})=-\mib{g}(\mib{k})$. 
The spin-orbit coupling lifts the two-fold degeneracy in the band as 
$\varepsilon_{\pm}(\mib{k})=\varepsilon(\mib{k})\pm\alpha|\mib{g}(\mib{k})|$. 
In this research, we study the two-dimensional Rashba spin-orbit coupling with 
$\mib{g}(\mib{k})=(-\sin{k_y},\sin{k_x},0)$ as well as the 
three-dimensional cubic spin-orbit coupling with 
$\mib{g}(\mib{k})=(\sin{k_x},\sin{k_y},\sin{k_z})$.

We take into account the mean field of superconducting order parameters in the last term 
of Eq.~(\ref{model}). 
The order parameters $\Delta_{ss'}(\mib{k})$ involve both spin singlet and triplet components
due to the spin-orbit coupling.
We here ignore the spin triplet component and assume the s-wave spin singlet order parameter 
[$\Delta_{\uparrow\downarrow}(\mib{k})=-\Delta_{\downarrow\uparrow}(\mib{k})=\psi$],
since the spin susceptibility at zero temperature is independent of 
the symmetry of order parameters for a large spin-orbit coupling $|\Delta_{ss'}(\mib{k})| \ll \alpha$.
~\cite{JPSJ.76.034712,PhysRevB.76.094516,JPSJ.76.124709} 
Although the $p$-wave superconducting state of Li$_2$Pt$_3$B has been indicated by several 
experimental results,~\cite{PhysRevLett.98.047002,PhysRevB.86.220502}
we do not touch this possibility since our analysis of the spin susceptibility 
cannot distinguish the $p$-wave superconducting state from the $s$-wave one. 
We take $|\psi|\leq 0.01$ so as to be small enough to satisfy the condition 
$|\Delta_{ss'}(\mib{k})|\ll\alpha$, as realized in most non-centrosymmetric superconductors.

\section{Spin Susceptibility in the Superconducting State}

In this section, we calculate the spin susceptibility in the superconducting state. 
We consider the zero temperature $T=0$ throughout this paper.
The spin susceptibility $\chi={\rm{lim}}_{H\rightarrow 0}\langle M\rangle/H$ is obtained by calculating the magnetization $\langle M\rangle$ in the field $\mib{H}$ and taking the limit 
$\mib{H} \rightarrow 0$. 
The Zeeman coupling term is introduced as
$H_{\rm{Z}}=-(g\mu_{\rm{B}}/2)\Sigma_{\mib{k},s,s'}\mib{H}\cdot\mib{\sigma}_{ss'}c^{\dag}_{\mib{k}s}c_{\mib{k}s'}$
where we assume $g=2$ and $\mu_{\rm{B}}$ is the Bohr magneton.
We first study the two-dimensional systems with Rashba spin-orbit coupling. 
Later, we will show the results for three-dimensional systems with cubic spin-orbit coupling. 
For two-dimensional systems, we focus on the spin susceptibility in the {\it ab}-plane, 
since that along the {\it c}-axis is not reduced by the superconductivity.~\cite{New.J.Phys.6.115}
On the other hand, the spin susceptibility is isotropic in the cubic system. 
We discuss the spin susceptibility normalized by the normal state value 
$\chi_{\rm s}/\chi_{\rm n}$, where $\chi_{\rm s}$ is the spin susceptibility in the superconducting 
state. 
The normal state value of spin susceptibility $\chi_{\rm n}$ is calculated at $T=0$ for 
$|\Delta_{ss'}(\mib{k})|=0$.

\subsection{Two-dimensional systems}

\begin{figure}[htbp]
  \begin{center}
    \includegraphics[scale=0.33]{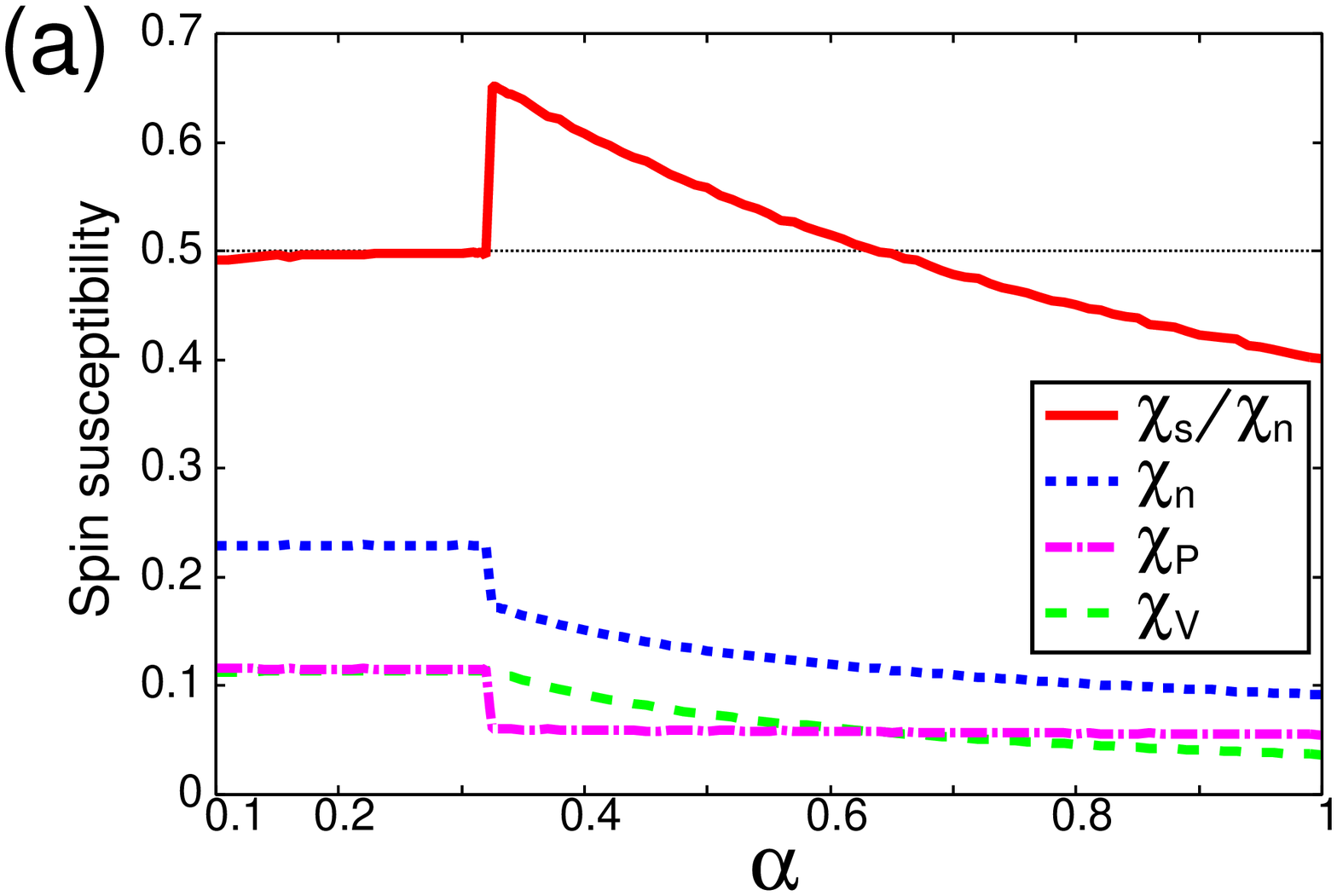}
    \includegraphics[scale=0.33]{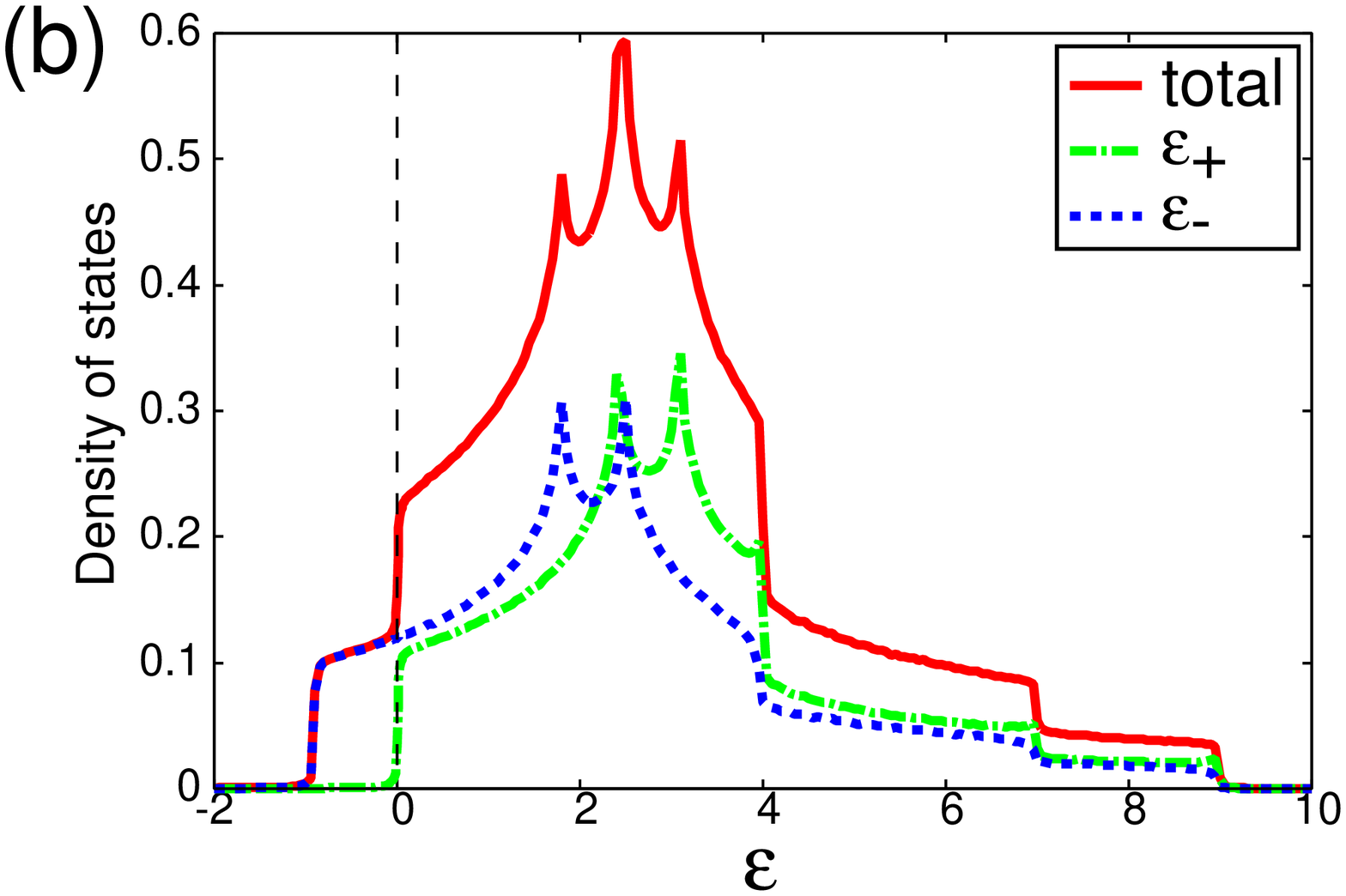}
  \end{center}
  \caption{(Color online)
    (a) Normalized spin susceptibility $\chi_{\rm s}/\chi_{\rm n}$ in the two-dimensional 
    systems with the Rashba spin-orbit coupling (solid line). 
    We assume the dispersion relation Eq.~(\ref{2D}) with $(t_1,t_2,t_3,n)=(-0.25,0.5,0.8,0.1)$. 
    The FS topological transition of type (A) occurs at $\alpha=0.325$. 
    The spin susceptibility in the normal state $\chi_{\rm n}$ is shown by the dotted line.  
    The Pauli part $\chi_{\rm P}$ (dash-dotted line) and Van-Vleck part $\chi_{\rm V}$ (dashed line) 
    are shown for discussion. 
    In Figs.~2(a), 3, and 4, the horizontal thin dotted line shows the normalized spin susceptibility 
    $\chi_{\rm s}/\chi_{\rm n}=1/2$ of conventional Rashba superconductors.~\cite{New.J.Phys.6.115} 
    (b) Density of states in the normal state at $\alpha=0.325$. 
    The solid line shows the total density of states. The density of states of the $\varepsilon_+$ band 
    and that of the $\varepsilon_-$ band are shown by the dash-dotted and dotted lines, 
    respectively. 
    The vertical dashed line shows the Fermi energy.
  }
  \label{2dtype}
\end{figure}

First, we study two-dimensional systems with the Rashba spin-orbit coupling 
$\mib{g}(\mib{k})=(-\sin{k_y},\sin{k_x},0)$. 
The dispersion relation is assumed as 
\begin{eqnarray}
\label{2D}
\varepsilon(\mib{k}) &=& 2t_1(\cos{k_x}+\cos{k_y})+4t_2\cos{k_x}\cos{k_y} \nonumber \\
&+& 2t_3(\cos{2k_x}+\cos{2k_y})-\mu.
\end{eqnarray}
The FS topological transition of type (A) occurs at $\alpha = 0.325$ 
for the parameters $(t_1,t_2,t_3,n)=(-0.25,0.5,0.8,0.1)$, 
where the band width is $W=8$. 
Figure~\ref{2dtype}(a) shows the normalized spin susceptibility 
$\chi_{\rm s}/\chi_{\rm n}$ as a function of the spin-orbit coupling $\alpha$. 
We see the discontinuous jump of the normalized spin susceptibility at $\alpha = 0.325$. 
This jump is caused by the sudden decrease in spin susceptibility in the normal state $\chi_{\rm n}$ 
at the type (A) FS topological transition with increasing $\alpha$. 
On the other hand, spin susceptibility in the superconducting state $\chi_{\rm s}$ is not 
substantially affected by the FS topological transition, as shown in Fig.~\ref{2dtype}(a).

We clarify these changes by dividing the spin susceptibility 
into the Pauli part and Van-Vleck part. 
The Van-Vleck part of spin susceptibility is defined using the dynamical spin susceptibility in 
the normal state $\chi_{\rm n}({\bm q},\omega)$ as 
$\chi_{\rm V} = {\rm lim}_{\omega \rightarrow 0} \ {\rm lim}_{{\bm q} \rightarrow 0} \ \chi_{\rm n}({\bm q},\omega)$. 
On the other hand, the spin susceptibility observed in experiments is obtained as 
$\chi_{\rm n} = {\rm lim}_{{\bm q} \rightarrow 0} \ {\rm lim}_{\omega \rightarrow 0} \ \chi_{\rm n}({\bm q},\omega)$.  
The Pauli part is the difference $\chi_{\rm P} = \chi_{\rm n} - \chi_{\rm V}$. 
The Pauli part $\chi_{\rm P}$ comes from the intraband contributions and is completely suppressed at $T=0$.
On the other hand, the Van-Vleck part $\chi_{\rm V}$ comes from the interband transition 
between the $\varepsilon_+$ band and the $\varepsilon_-$ band 
and is hardly affected by the superconductivity. 
Note that this Van-Vleck part $\chi_{\rm V}$ is different from 
the usual $T$-independent Van-Vleck susceptibility arising from the orbital degrees of freedom.
The Van-Vleck part $\chi_{\rm V}$ in our definition has a temperature dependence similarly 
to the Pauli part $\chi_{\rm P}$ when the spin-orbit coupling $\alpha$ is much smaller 
than the Fermi energy. 
Thus, this $\chi_{\rm V}$ is included in the spin part of the NMR Knight shift $K_{\rm s}$, 
while the Van-Vleck susceptibility arising from the orbital degrees of freedom is included 
in the orbital part $K_{\rm orb}$. 
When we focus on the spin susceptibility extracted from the spin part $K_{\rm s}$, 
as often analyzed in the NMR experiment, the spin susceptibility is obtained 
as $\chi_{\rm n} = \chi_{\rm P} + \chi_{\rm V}$ in the normal state, while it is almost 
equivalent to the Van-Vleck part $\chi_{\rm s} \approx \chi_{\rm V}$ in the superconducting state.

Because the Pauli part $\chi_{\rm P}$ is proportional to the density of states at the Fermi level, 
the disappearance of the Fermi surface at the type (A) FS topological transition decreases 
$\chi_{\rm P}$ as well as $\chi_{\rm n}$. This decrease occurs in a discontinuous manner in the 
two-dimensional systems since the density of states is discontinuous at the band edge 
[see Fig.~\ref{2dtype}(b)]. 
On the other hand, the spin susceptibility in the superconducting state is robust 
for the disappearance of the Fermi surface, since that comes from the Van-Vleck term 
$\chi_{\rm V}$. 
In this way, the normalized spin susceptibility $\chi_{\rm s}/\chi_{\rm n}$ is increased 
at the type (A) FS topological transition with increasing the spin-orbit coupling. 
For a large spin-orbit coupling $\alpha >0.325$, $\chi_{\rm s}/\chi_{\rm n}$ gradually decreases 
with $\alpha$, because of the decrease in the Van-Vleck term. 
It should be stressed that $\chi_{\rm s}/\chi_{\rm n}$ is much larger than the canonical value $1/2$ 
when the spin-orbit coupling $\alpha$ is a little larger than the critical value 
$\alpha_{\rm c} = 0.325$.

\begin{figure}[tbp]
  \begin{center}
    \includegraphics[scale=0.33]{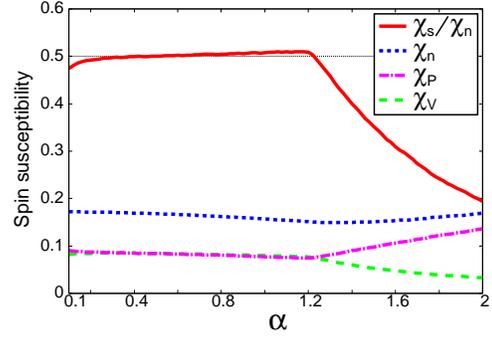}
  \end{center}
  \caption{(Color online)
    Normalized spin susceptibility $\chi_{\rm s}/\chi_{\rm n}$ for $(t_1,t_2,t_3,n)=(-1,0,0,0.1)$ 
in Eq.~(\ref{2D}). 
The FS topological transition of type (B) occurs at $\alpha =1.22$. 
    The lines show the same quantities as in Fig.~2(a). 
  }
  \label{2ddirac}
\end{figure}

Next, we study the FS topological transition of type (B).
For the parameters $(t_1,t_2,t_3,n)=(-1,0,0,0.1)$ of Eq.~(\ref{2D}), 
the Fermi surface crosses the Dirac point at $\alpha=1.22$. 
Figure~\ref{2ddirac} shows the decrease in $\chi_{\rm s}/\chi_{\rm n}$ for $\alpha>1.22$
in sharp contrast to the FS topological transition of type (A). 
This is because the density of states at the Fermi energy increases and therefore the Pauli part  
$\chi_{\rm P}$ increases with $\alpha$ for $\alpha > 1.22$.

\begin{figure}[tbp]
  \begin{center}
    \includegraphics[scale=0.33]{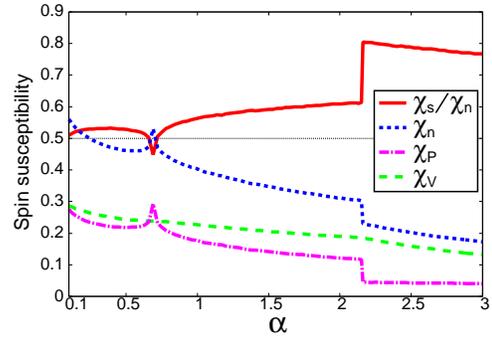}
  \end{center}
  \caption{(Color online)
    Normalized spin susceptibility $\chi_{\rm s}/\chi_{\rm n}$ for 
$(t_1,t_2,t_3,n)=(-0.25,0.5,0.8,0.8)$ 
    in Eq.~(\ref{2D}). 
    The FS topological transition of type (C) occurs at $\alpha =0.69$, and 
    that of type (A) occurs at $\alpha=2.16$. 
    The lines show the same quantities as in Fig.~2(a). 
  }
  \label{2dmax}
\end{figure}

The normalized spin susceptibility $\chi_{\rm s}/\chi_{\rm n}$ is increased by the FS topological transition 
of type (C), as investigated by Fujimoto.~\cite{JPSJ.76.034712}
Our calculation reproduces his result, but the enhancement of normalized spin susceptibility 
is much smaller than that due to the type (A) FS topological transition. 
When we assume the parameters $(t_1,t_2,t_3,n)=(-0.25,0.5,0.8,0.8)$ of Eq.~(\ref{2D}), 
the FS topological transition of type (C) occurs at $\alpha =0.69$.  
Figure~\ref{2dmax} shows that $\chi_{\rm s}/\chi_{\rm n}$ decreases at the transition $\alpha =0.69$ 
and increases with increasing $\alpha$ for $\alpha > 0.69$. 
The increase in $\chi_{\rm s}/\chi_{\rm n}$ for $\alpha > 0.69$ 
is less pronounced than that due to the type (A) transition. 
Indeed, we see a significant increase in $\chi_{\rm s}/\chi_{\rm n}$ 
at $\alpha=2.16$ where the FS topological transition of type (A) occurs. 

\subsection{Three-dimensional systems}

\begin{figure}[bp]
  \begin{center}
    \includegraphics[scale=0.33]{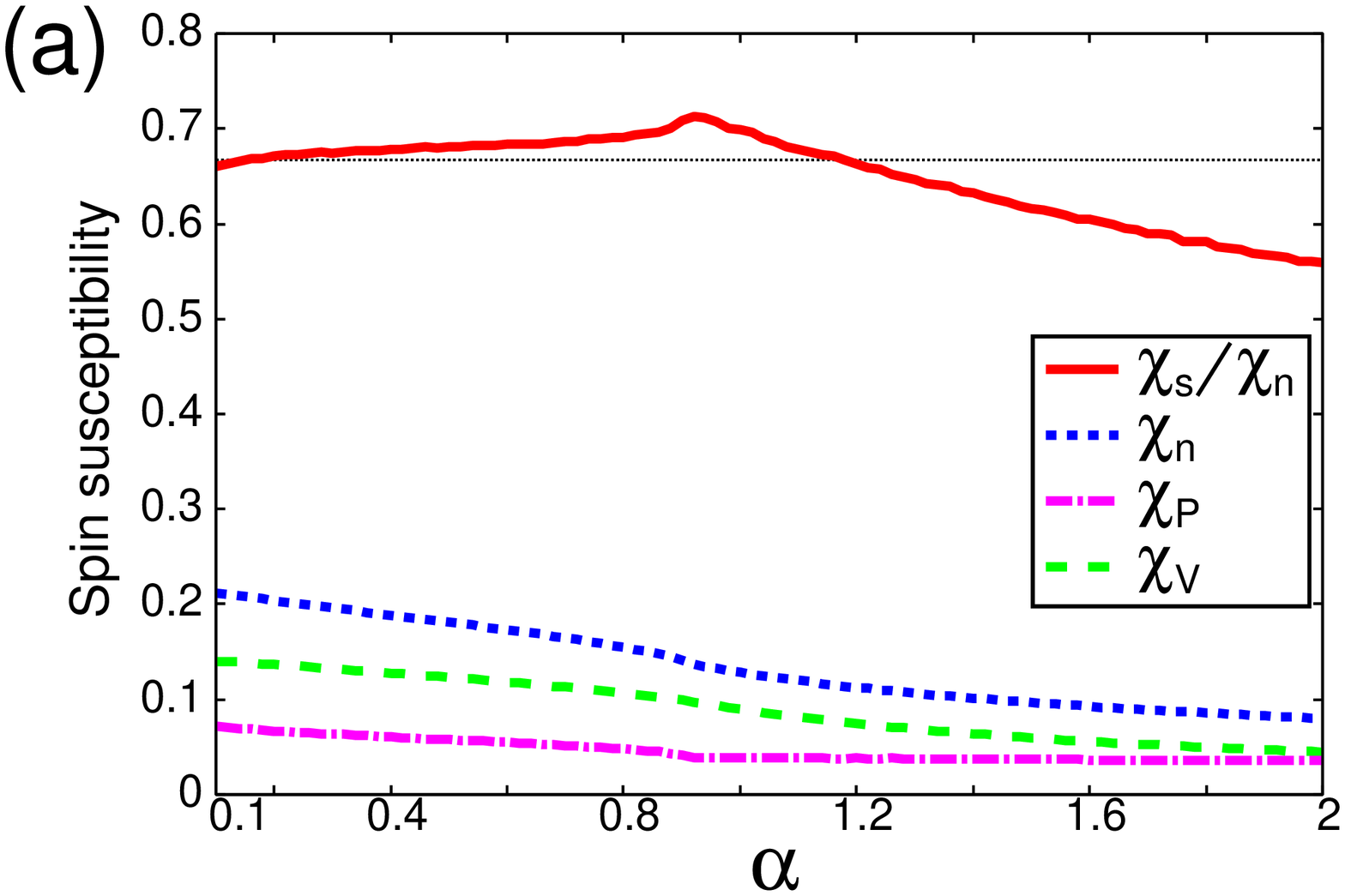}
    \includegraphics[scale=0.33]{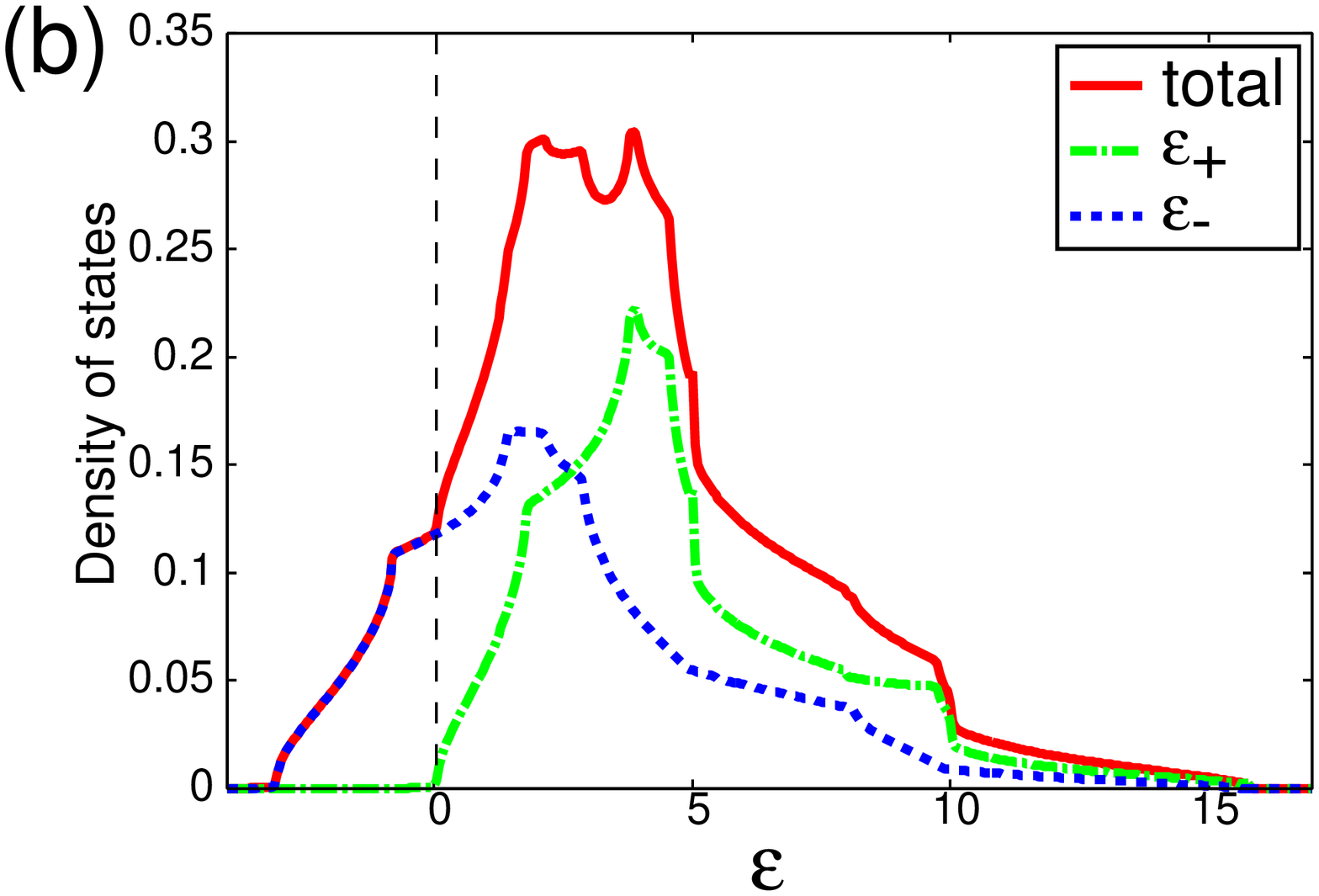}
  \end{center}
  \caption{(Color online)
    (a) Normalized spin susceptibility $\chi_{\rm s}/\chi_{\rm n}$ in the three-dimensional systems 
    with the cubic spin-orbit coupling. 
    We assume the parameters 
    $(t_1,t_2,t_3,t_4,n)=(-0.8,0.275,0.1125,0.8,0.2)$ in Eq.~(\ref{3D}). 
    The FS topological transition of type (A) occurs at $\alpha=0.92$. 
    The lines show the same quantities as in Fig.~2(a). 
In Figs.~5(a) and 6, the horizontal thin dotted line shows the normalized spin susceptibility 
      $\chi_{\rm s}/\chi_{\rm n}=2/3$ of conventional cubic non-centrosymmetric superconductors.~\cite{JPSJ.76.034712,PhysRevB.76.094516} 
    (b) Density of states in the normal state at $\alpha=0.92$.
    The vertical dashed line shows the Fermi energy.
    The lines show the same quantities as in Fig.~2(b). 
  }
  \label{3dtype}
\end{figure}

We turn to three-dimensional systems with the cubic symmetry.  
The cubic spin-orbit coupling with $\mib{g}(\mib{k})=(\sin{k_x},\sin{k_y},\sin{k_z})$ is considered here. 
We assume the dispersion relation as,
\begin{eqnarray}
\label{3D}
\varepsilon(\mib{k})&=&2t_1(\cos{k_x}+\cos{k_y}+\cos{k_z}) \nonumber \\
&+&4t_2(\cos{k_x}\cos{k_y}+\cos{k_y}\cos{k_z}+\cos{k_z}\cos{k_x}) \nonumber \\
&+&8t_3\cos{k_x}\cos{k_y}\cos{k_z} \nonumber \\
&+&2t_4(\cos{2k_x}+\cos{2k_y}+\cos{2k_z})-\mu.
\end{eqnarray}
When we choose the parameters 
$(t_1,t_2,t_3,t_4,n)=(-0.8,0.275,0.1125,0.8,0.2)$, where the band width is $W=17.24$, 
the Fermi surfaces show the topological transition 
of type (A) at $\alpha=0.92$. 
Figure~\ref{3dtype}(a) shows the maximum $\chi_{\rm s}/\chi_{\rm n}$ at this FS topological transition.  
On the other hand, the increase in $\chi_{\rm s}/\chi_{\rm n}$ is not discontinuous in contrast to 
that in two-dimensional systems [see Fig.~2(a)]. 
This is because the density of states continuously decreases as 
$\rho(\varepsilon) \propto \sqrt{\varepsilon- \varepsilon_{\rm c}}$ at the band edge 
$\varepsilon = \varepsilon_{\rm c}$ [see Fig.~\ref{3dtype}(b)]. 
Because of the less singular properties in the density of states, the increase in the normalized 
spin susceptibility $\chi_{\rm s}/\chi_{\rm n}$ due to the FS topological transition is less 
pronounced than that in two-dimensional systems.

\begin{figure}[htbp]
  \begin{center}
    \includegraphics[scale=0.33]{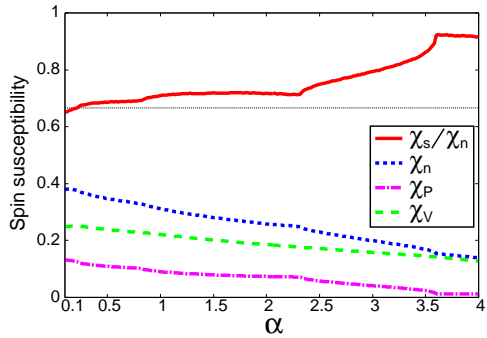}
  \end{center}
  \caption{(Color online)
Normalized spin susceptibility $\chi_{\rm s}/\chi_{\rm n}$ for 
$(t_1,t_2,t_3,t_4,n)=(-0.8,0.275,0.1125,0.8,0.9)$ in Eq.~(\ref{3D}). 
The FS topological transition of type (C) occurs at $\alpha=0.82$ and $\alpha =2.3$, 
and that of type (A) occurs at $\alpha =3.6$. 
  }
  \label{3dmax}
\end{figure}

A small enhancement of $\chi_{\rm s}/\chi_{\rm n}$ in Fig.~5(a) 
from the conventional value $\chi_{\rm s}/\chi_{\rm n} = 2/3$
implies that there is another source of the large spin susceptibility 
$\chi_{\rm s}/\chi_{\rm n}$ observed in Li$_2$Pt$_3$B. 
We here show a case in which a large $\chi_{\rm s}/\chi_{\rm n}$ close to unity 
is realized. 
When we assume the parameters $(t_1,t_2,t_3,t_4,n)=(-0.8,0.275,0.1125,0.8,0.9)$ 
in Eq.~(\ref{3D}), the  Fermi surface of the $\varepsilon_-$ band and that of 
the $\varepsilon_+$ band cross the van-Hove singularity at $\alpha=0.82$ and $\alpha =2.3$, 
respectively. With further increase in the spin-orbit coupling, 
the Fermi surface of the $\varepsilon_+$ band vanishes at $\alpha =3.6$. 
We obtain a large normalized spin susceptibility $\chi_{\rm s}/\chi_{\rm n} >0.9$ 
for $\alpha >3.6$, as shown in Fig.~\ref{3dmax}. 

We explain such a large spin susceptibility in the superconducting state by discussing again 
the Pauli part and Van-Vleck part of spin susceptibility. 
For simplicity, we consider a small electron pocket Fermi surface of heavy $\varepsilon_+$ band and 
a small hole pocket of light $\varepsilon_-$ band. The Pauli part of spin susceptibility 
is proportional to the density of states,
\begin{eqnarray}  
\rho(\epsilon_{\rm F}) = \frac{\sqrt{2m_{\rm e}^3 (\epsilon_{\rm F} - \epsilon_{\rm c+})}}{4\pi^2} + 
\frac{\sqrt{2m_{\rm h}^3 (\epsilon_{\rm c-} - \epsilon_{\rm F})}}{4\pi^2}, \nonumber
\end{eqnarray}
where the first term (second term) 
comes from the heavy electron band (light hole band).  
We denoted the band edge of each band, $\epsilon_{\rm c+}$ and $\epsilon_{\rm c-}$, and assume 
the effective mass, $m_{\rm e} \gg m_{\rm h}$.  
When the electron pocket vanishes with increasing antisymmetric spin-orbit coupling, 
the density of states is significantly decreased as 
\begin{eqnarray}
\rho(\epsilon_{\rm F}) = \frac{\sqrt{2m_{\rm h}^3 (\epsilon_{\rm c-} - \epsilon_{\rm F})}}{4\pi^2}. \nonumber
\end{eqnarray} 
The decrease in the density of states leads to a decrease in the Pauli part spin susceptibility, 
while the Van-Vleck part is hardly affected. 
Thus, the normalized spin susceptibility $\chi_{\rm s}/\chi_{\rm n} = 
\chi_{\rm V}/(\chi_{\rm P} + \chi_{\rm V})$ shows a substantial increase as it approaches 
the FS topological transition of type (A), when the spin-split Fermi surfaces 
have different effective masses and the Fermi surface of heavy band vanishes, as in the case of our model adopted in Fig.~\ref{3dmax}.
This is a possible mechanism of the large spin susceptibility $\chi_{\rm s}/\chi_{\rm n}$ 
in Li$_2$Pt$_3$B, although the possibility of another source for realizing a small Pauli term is not excluded. 
We would like to stress that such a small Pauli term 
is not realized by a small antisymmetric spin-orbit coupling compared with the Fermi energy.

Indeed, successive FS topological transitions from  Li$_2$Pd$_3$B to Li$_2$Pt$_3$B 
have been indicated by the first-principles band structure 
calculations.~\cite{PhysRevB.72.174505,Shishidou}
Thus, the intriguing topology of the Fermi surface in Li$_2$(Pd$_{1-x}$Pt$_{x}$)$_3$B may be the source of 
the unusual magnetic properties in the superconducting state. 
A decrease in the density of states with increasing concentration of Pt ions has not been clearly 
observed~\cite{PhysRevB.76.104506}, indicating that our proposal is not likely realized. 
However, the multiband structure of Li$_2$(Pd$_{1-x}$Pt$_{x}$)$_3$B does not allow such a simple 
discussion.
FS topological transitions of Li$_2$(Pd$_{1-x}$Pt$_{x}$)$_3$B 
are partly due to the multiband structure, and the single-band model 
adopted in this paper does not precisely reproduce the electronic structure of 
Li$_2$(Pd$_{1-x}$Pt$_{x}$)$_3$B. 
The analysis of a realistic model is desired to elucidate 
the superconducting state of Li$_2$(Pd$_{1-x}$Pt$_{x}$)$_3$B. 

\section{Summary and Discussion}

We have investigated the spin susceptibility of non-centrosymmetric superconductors, which is accompanied 
by the topological transition of Fermi surfaces owing to the antisymmetric spin-orbit coupling. 
When one of the Fermi surfaces of the spin-split band vanishes [FS topological transition of type (A)], 
the normalized spin susceptibility $\chi_{\rm s}/\chi_{\rm n}$ is increased. 
On the other hand, $\chi_{\rm s}/\chi_{\rm n}$ is decreased by the FS topological transition of type (B) 
in which the Fermi level crosses the Dirac point. 
The spin susceptibility $\chi_{\rm s}/\chi_{\rm n}$ increases when the Fermi surface crosses van-Hove singularities 
[FS topological transition of type (C)], but the increase is smaller than that due to type (A). 
We obtain the maximum $\chi_{\rm s}/\chi_{\rm n}$ at the type (A) 
FS topological transition, and a large $\chi_{\rm s}/\chi_{\rm n}$ 
for the antisymmetric spin-orbit coupling $\alpha$ larger than the critical value.

These behaviors of the spin susceptibility are understood in terms of the density of states. 
The density of states depends on the band structure; however, it shows a universal change 
at the FS topological transition. Thus, our results on the changes at the FS topological transition 
are qualitatively independent of the band structure. 
Note that these results are also independent of the symmetry of superconductivity.

The effects of FS topological transitions are pronounced in the two-dimensional systems 
because of the discontinuous jump of the density of states at the band edge. 
Even in three-dimensional systems, the spin susceptibility is almost unchanged through the superconducting 
transition, when successive transitions of types (A) and (C) occur. 
We obtained a large normalized spin susceptibility 
$\chi_{\rm s}/\chi_{\rm n} >0.9$, which is consistent with the NMR Knight shift measurement for 
Li$_2$Pt$_3$B~\cite{PhysRevLett.98.047002,PhysRevB.86.220502} within the experimental resolution. 
Generally, such a large normalized spin susceptibility is obtained when 
the density of states is significantly decreased by the antisymmetric spin-orbit coupling. 
We showed an example of such a band structure. 
Although our single-band model does not reproduce the multiband structure of Li$_2$Pt$_3$B, 
our finding indicates the important roles of FS topological transitions. 
Indeed, the band structure calculation shows a lot of topological transitions 
in Li$_2$Pt$_3$B owing to the large spin-orbit coupling, but not in 
Li$_2$Pd$_3$B because of the small spin-orbit coupling~\cite{PhysRevB.72.174505,Shishidou}. 
In order to elucidate the effect of the intriguing topology of the Fermi surface on the 
superconducting phase in Li$_2$Pt$_3$B, it is desired to study the multiorbital model, 
which precisely describes the electronic structure of Li$_2$(Pd$_{1-x}$Pt$_{x}$)$_3$B.

\section*{Acknowledgements}
The authors are grateful to T. Shishidou and G.-q. Zheng for fruitful discussions. 
This work was supported by a Grant-in-Aid for Scientific Research 
on Innovative Areas ``Heavy Electrons'' (No. 23102709) from MEXT, 
and by a Grant-in-Aid for Young Scientists (B) (No. 24740230) from JSPS.

\end{document}